\documentclass[aps,prl,twocolumn,groupedaddress,showpacs,amsmath,amssymb]{revtex4}

\usepackage{graphicx}
\usepackage{dcolumn}
\usepackage{bm}

\begin{document}

\title{Spin-Orbit Coupling and Anomalous Angular-Dependent 
Magnetoresistance in the Quantum Transport Regime of PbS}

\author{Kazuma Eto}
\author{A. A. Taskin}
\author{Kouji Segawa}
\author{Yoichi Ando}
\affiliation{Institute of Scientific and Industrial Research, 
Osaka University, Ibaraki, Osaka 567-0047, Japan}

%\date{\today}

\begin{abstract}

We measured magnetotransport properties of PbS single crystals which
exhibit the quantum linear magnetoresistance (MR) as well as the
static skin effect that creates a surface layer of additional
conductivity. The Shubnikov-de Haas oscillations in the longitudinal MR
signify the peculiar role of spin-orbit coupling. In the
angular-dependent MR, sharp peaks are observed when the magnetic field
is slightly inclined from the longitudinal configuration, which is
totally unexpected for a system with nearly spherical Fermi surface and
points to an intricate interplay between the spin-orbit coupling and the
conducting surface layer in the quantum transport regime.

\end{abstract}


% 71.18.+y 	Fermi surface: calculations and measurements; effective mass, 
%               g factor
% 73.25.+i 	Surface conductivity and carrier phenomena
% 73.20.-r 	Electron states at surfaces and interfaces
% 73.20.At 	Surface states, band structure, electron density of states 
% 71.70.Di 	Landau levels
%
% 72.20.-i      Conductivity phenomena in semiconductors and insulators  
% 72.20.My         Galvanomagnetic and other magnetotransport effects 
% 72.20.Ht         High-field and nonlinear effects 
%
% 75.47.-m      Magnetotransport phenomena; materials for magnetotransport  

\maketitle

Recently, non-trivial consequences of the spin-orbit coupling (SOC) in
crystalline solids are a major theme in condensed matter physics
\cite{Nagaosa}. For example, the spin Hall effect is a striking
manifestation of the SOC in non-magnetic materials \cite{Murakami}, and
the SOC in non-centrosymmetric superconductors gives rise to an
unconventional order-parameter symmetry \cite{Fujimoto}. Even more
strikingly, it was recognized that a certain class of narrow-gap
semiconductors where the energy gap is a product of the SOC are
topological insulators, whose valence band structures is characterized
by a non-trivial $Z_2$ topological invariant
\cite{K3,SCZ2,MB,K1,K2,SCZ1}. The three-dimensional topological
insulators host helically spin-polarized surface states and are
predicted to exhibit various novel phenomena
\cite{K1,K2,SCZ1,Majorana,Monopole}. After the discovery of the
topological-insulator nature in Bi$_{1-x}$Sb$_x$, Bi$_2$Se$_3$, and
Bi$_2$Te$_3$ \cite{H1,SCZ4,H5,Shen}, those three materials are under
intense investigations.

In this context, the narrow-gap semiconductor PbS would make a useful
comparison, because its energy gap is due to a strong SOC but its
valence band structure lends itself to the trivial $Z_2$ topological
class \cite{K2}; namely, PbS is a non-topological insulator.
Nevertheless, this material may be called an ``incipient" spin Hall
insulator, since the energy gap of the SOC origin in PbS causes a large
Berry phase in the Bloch states and leads to a finite intrinsic spin
Hall conductivity $\sigma_{\rm H}^{\rm s}$ even in the insulating state
\cite{Mura2}. Therefore, the role of the SOC in the transport properties
of this material is worth investigating with the modern understanding. 

PbS has a rock salt crystal structure and has a direct energy gap of
about 0.3 eV located at the four equivalent $L$ points of the Brillouin
zone \cite{Dalven}. Depending on whether S is excessive or deficient,
both $p$- and $n$-type PbS can be prepared, and in both cases the Fermi
surface (FS) is very nearly spherical \cite{Dalven,Stiles}. This
material was well studied in the past for its potential in the infrared
applications \cite{Dalven}. More recently, PbS is attracting attentions
in the photovoltaic community because of the multi-exiton generation
\cite{Pijpers}. In this Letter, we report our detailed study of the
magnetoresistance (MR) in low-carrier-density PbS, focusing on its
angular dependence. To our surprise, we observed sharp peaks in the
angular-dependent MR in high magnetic fields, which is totally
unexpected for a three-dimensional (3D) material with a small spherical
FS. Although the exact mechanism of this anomalous behavior is not clear
at the moment, our data points to an important role of the SOC in the
quantum transport regime. In addition, the formation of a surface layer
with additional conductivity due to skipping orbits (called ``static
skin effect" \cite{Azbel}) appears to be also playing a role in the
observed angular dependence. The unexpected angular-dependent MR points
to a hitherto-overlooked effect that could become important in the
magnetotransport properties of narrow-gap semiconductors with a strong
SOC.

\begin{figure}
\includegraphics*[width=8.0cm]{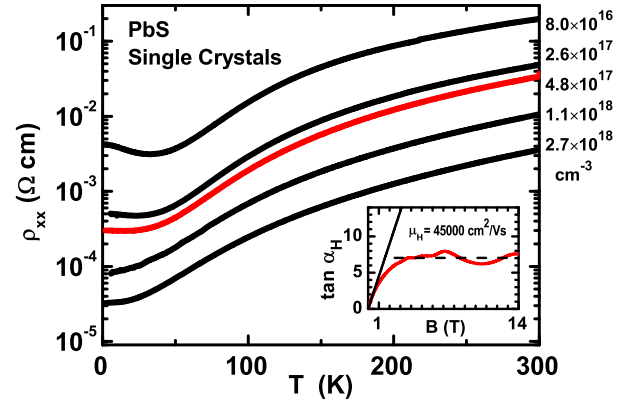}
\caption{(Color online) 
Temperature dependence of $\rho_{xx}$ for a series of $p$-type 
PbS single crystals, whose $n_h$ is indicated on the right. 
Inset shows $\tan \alpha_{\rm H}$ vs $B$
for the 4.8$\times$10$^{17}$ cm$^{-3}$ sample,
which exhibits a deviation from the classical linear behavior 
(shown by the solid straight line) and a saturation above 
4 T in the quantum transport regime; the low-field slope is equal 
to $\mu_{\rm H}$ and gives 4.5$\times$10$^{4}$ cm$^2$/Vs. 
}
\label{fig1}
\end{figure}

High-quality single crystals of PbS were grown by a vapor transport
method from a stoichiometric mixture of 99.998\% purity Pb and 99.99\%
purity S. The mixture was sealed in an evacuated quartz tube and was
reacted for 5 -- 10 h at 980$^{\circ}$C. After the reaction, the
resulting material was vaporized and transported to the other end of the
sealed tube by making a large temperature difference, which worked as a
purification stage. The obtained polycrystals were taken out and again
sealed in a new evacuated quartz tube for the crystal growth stage: The
polycrystal-containing end of the tube was kept at 850$^{\circ}$C, and
sublimed PbS was transported to the other end kept at 840$^{\circ}$C for
one week. The obtained single crystals were annealed in sulfur vapor to
tune the type and the density of carriers, during which the crystal
temperature and the sulfur vapor pressure were controlled independently.
We have prepared a series of samples with various carrier density as
shown in Fig. 1. In the following we focus on a $p$-type sample with the
carrier density $n_h$ of 4.8$\times$10$^{17}$ cm$^{-3}$, which was
obtained by annealing the crystal at 533$^{\circ}$C with the sulfur
vapor source kept at 90$^{\circ}$C. 

The resistivity $\rho_{xx}$ and the Hall resistivity $\rho_{yx}$ were
measured simultaneously by using a standard six-probe method on a thin
rectangular sample whose top and bottom surfaces were cleaved (001)
plane. The current $I$ was always along the [100] direction. The
Shubnikov-de Haas (SdH) oscillations were measured by sweeping the
magnetic field $B$ between +14 and -14 T for a series of field
directions. Continuous rotations of the sample in constant magnetic
fields were used for measuring the angular dependence of the MR, in
which the direction of the magnetic field $B$ was either [001]
$\rightarrow$ [010] (transverse geometry) or [001] $\rightarrow$ [100]
(transverse to longitudinal geometry).

\begin{figure}
\includegraphics*[width=7.0cm]{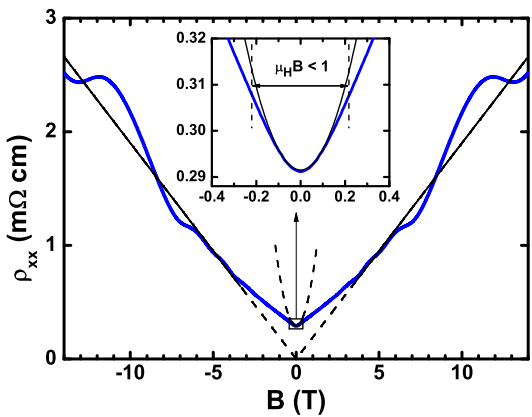}
\caption{(Color online) 
Transverse MR at 1.5 K with $B$ along [001]. 
The straight lines are the background linear MR, 
on top of which pronounced SdH oscillations are superimposed. 
Inset shows the $B^2$ dependence observed in the low-field
classical regime ($\mu_{\rm H}B <$ 1) and its fitting (thin solid line).
}
\label{fig2}
\end{figure}

Figure 2 shows the transverse MR measured at 1.5 K with $B$ along [001].
Pronounced SdH oscillations are clearly seen, and one may notice that
the background MR does not show the ordinary $B^2$ dependence. This is
because the range of the weak-field regime ($\mu_{\rm H}B <$ 1), where
the $B^2$ dependence is observed, is extremely narrow in our sample, as
shown in the inset of Fig. 2 ($\mu_{\rm H}$ is the Hall mobility). It is
worth noting that our sample shows a nearly-$B$-linear background MR in
the strong-field regime ($\mu_{\rm H}B >$ 1), rather than a tendency to
saturation which is usually observed in metals. This high-field behavior
is the so-called ``quantum linear MR" proposed by Abrikosov
\cite{Abrikosov}. Such a behavior is expected in the quantum transport
regime where the condition $n_h < (eB/\hbar c)^{2/3}$ is satisfied
\cite{Abrikosov}, and in our sample this regime is realized for $B >$ 4
T. In this sense, our PbS presents a straightforward realization of the
quantum linear MR and is different from Ag$_{2+\delta}$Se or
Ag$_{2+\delta}$Te where the linear MR is observed down to very low field
\cite{Abrikosov}.

\begin{figure}
\includegraphics*[width=8.0cm]{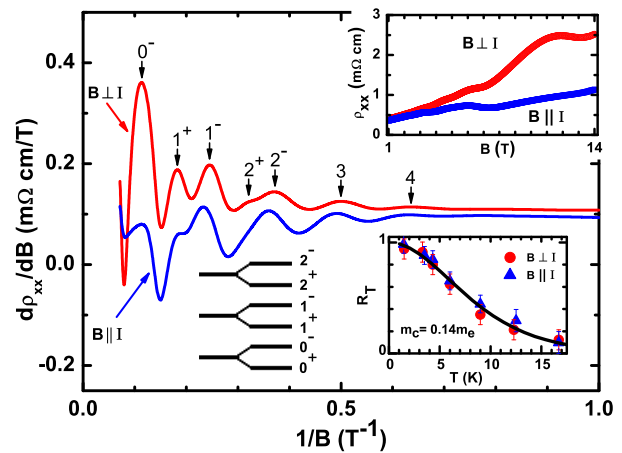}
\caption{(Color online) 
SdH oscillations in the transverse ($B \perp I$) and longitudinal ($B
\parallel I$) MR at 1.5 K. Upper inset shows the raw data, and the main
panel presents $d\rho_{xx}/dB$ vs $1/B$. In the ``$B \parallel I$"
measurement, the magnetic field was 8$^{\circ}$ off from the exactly
parallel direction. Lower left inset shows the schematic Landau level
diagram. Lower right inset shows the temperature dependence of the
normalized SdH amplitude, which yields $m_c$ = 0.14$m_e$ ($m_e$ is the
free electron mass).
}
\end{figure}
\label{fig3}

The SdH oscillations measured in the transverse ($B \perp I$) and
longitudinal ($B \parallel I$) MR are presented in Fig. 3 by plotting
$d\rho_{xx}/dB$ vs $1/B$. One can see that we are resolving the
spin-splitting of the Landau levels in high magnetic fields and that the
crossing of the 0$^-$ state by the Fermi level is observed (see the
lower left inset of Fig. 3 for the Landau level diagram); this means
that all the electrons are in the 0$^+$ state in the highest field (14
T) and hence the system is in the quantum limit. Also, one may notice
that the amplitude of some particular peaks, 0$^-$, 1$^+$, and 2$^+$,
are significantly diminished in the longitudinal configuration ($B
\parallel I$) compared to the transverse one where all the peaks are
well developed. Such a behavior was previously observed in
Hg$_{1-x}$Cd$_x$Te \cite{Narita1} and in Pb$_{1- x}$Sn$_x$Te
\cite{Narita2}, and was elucidated to be due to some selection rules
\cite{Narita1} imposed by the SOC which prohibits scattering between
certain Landau sublevels in the longitudinal configuration. Besides this
peculiar difference, the peak positions are almost the same for the two
field directions, which is because the FS in PbS is nearly spherical
\cite{Dalven,Stiles}. Also, the cyclotron mass $m_c$ extracted from the
temperature dependence of the SdH amplitude (lower right inset) using
the Lifshitz-Kosevich formula \cite{Shnbrg} is identical for the two
directions. 

\begin{figure}
\includegraphics*[width=7.0cm]{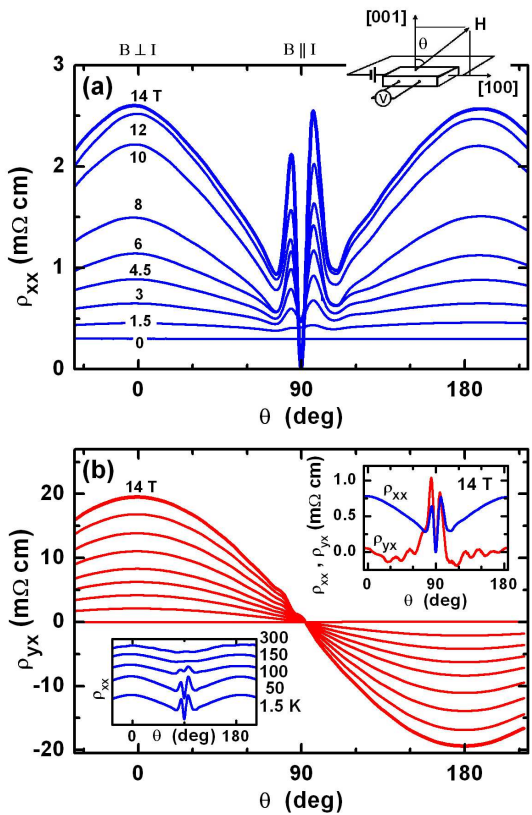}
\caption{(Color online) 
Angular-dependence of (a) $\rho_{xx}$ and (b) $\rho_{yx}$ for the
transverse-to-longitudinal rotation (top inset shows the geometry). In
panel (b), the upper inset compares the deviation of $\rho_{yx}(\theta)$
from the $\cos \theta$ dependence to $\rho_{xx}(\theta)$ multiplied by
0.3, while the lower inset shows how the anomalous peaks weaken with
temperature (data are vertically shifted for clarity).
}
\label{fig4}
\end{figure}

The carrier density $n_h$ calculated from the FS volume seen by the SdH
oscillations is 4.8$\times$10$^{17}$ cm$^{-3}$, which agrees well with
the value of $n_h$ = 4.6$\times$10$^{17}$ cm$^{-3}$ obtained from the
high-field Hall coefficient $R_{H\infty}$. It is worth noting that the
Hall mobility $\mu_{\rm H}$ calculated from $R_{H\infty}$ and
$\rho_{xx}$ at 0 T is 4.5$\times$10$^{4}$ cm$^2$/Vs, while the mobility
$\mu_{\rm SdH}$ obtained from the SdH oscillations is only
3.8$\times$10$^{3}$ cm$^2$/Vs \cite{note1}. This discrepancy is
essentially due to the fact that $\rho_{xx}$ and the Dingle temperature
$T_D$ (smearing factor in the SdH effect \cite{Shnbrg}) are determined
by different scattering processes; namely, $\rho_{xx}$ is primarily
determined by the backward scattering, while $T_D$ is sensitive to both
forward and backward scattering \cite{Shnbrg,DasSarma}. Apparently, only
small-angle scatterings are relevant in high-mobility PbS, which leads
to the 12 times difference between $\mu_{\rm H}$ and $\mu_{\rm SdH}$.

Now let us present the most surprising result. The angular-dependent MR
for the transverse-to-longitudinal rotation ($I$ was along [100] and $B$
was rotated from [001] toward [100]) is shown in Fig. 4(a), and the
corresponding angular dependence of $\rho_{yx}$ is shown in Fig. 4(b)
(the magnetic-field angle $\theta$ is measured from [001]). In Fig.
4(a), pronounced peaks are observed when $\theta$ is near 90$^{\circ}$,
that is, when the magnetic field is slightly inclined from the
longitudinal configuration. The $\theta$ dependence of $\rho_{yx}$ also
shows a feature near 90$^{\circ}$, which can be more easily seen in the
upper inset of Fig. 4(b) where the deviation of the measured
$\rho_{yx}(\theta)$ from a smooth $\cos \theta$ dependence is plotted
together with the $\rho_{xx}(\theta)$ data (which is multiplied by 0.3).
One can easily see in this inset that the sharp peak occurs in both
$\rho_{xx}$ and $\rho_{yx}$ at the same $\theta$; in addition,
$\rho_{yx}(\theta)$ apparently shows periodic oscillations in a wide
range of $\theta$, whose relation to the sharp peak is not obvious. In
any case, given that the FS in PbS is nearly spherical and that they
cannot give rise to any open orbit, such a sharp peak in the
angular-dependent MR is totally unexpected. 

\begin{figure}\includegraphics*[width=8.5cm]{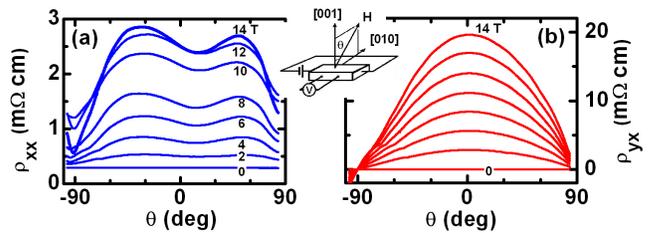}
\caption{(Color online) 
Angular-dependence of (a) $\rho_{xx}$ and (b) $\rho_{yx}$ for the 
transverse rotation (inset shows the geometry). The $\rho_{xx}$ data 
shown here are after removing the admixture of the $\rho_{yx}$ 
component in the $\rho_{xx}$ measurement.
}
\label{fig5}
\end{figure}

To gain insight into the origin of the unexpected peak in
$\rho_{xx}(\theta)$, the angular-dependent MR data in a different
rotation plane is useful. Figure 5 shows such data for the transverse
rotation ($I$ was along [100] and $B$ was rotated from [001] toward
[010]). As one can see in Fig. 5, there is no sharp peak in this
configuration, which immediately indicates that the unexpected peaks are
peculiar to the near-longitudinal configuration. Besides the absence of
the sharp peaks, there is a notable feature in Fig. 5: Since PbS has a
cubic symmetry, the MR for $\theta$ = 0$^{\circ}$ ($B$ along [001]) and
90$^{\circ}$ ($B$ along [010]) should be the same, since [001] and [010]
are crystallographically identical and the measurement configuration is
both transverse. However, the actual data in Fig. 5 indicates that they
are different, which suggests that there must be some additional factor
which affects the resistivity in magnetic field. In the past, similar
anisotropy was observed in clean specimens of low-carrier-density
materials such as Bi \cite{Bogod1} and Sb \cite{Bogod2}, and was
explained in terms of the static skin effect \cite{Azbel}, that is, the
formation of a surface layer of additional conductivity due to skipping
orbits when the magnetic field is nearly parallel to a specular surface.
In this regard, since the top and bottom surfaces of our sample were
cleaved (001) plane, it is understandable that the static skin effect
creates a surface conduction layer near the specular (001) surface for
the magnetic field along [010], leading to a reduced resistivity. 

An additional factor to consider regarding the MR anisotropy in the
present case is the SOC which diminishes some of the peaks in the SdH
oscillations for $B \parallel I$. In fact, as one can infer in the upper
inset of Fig. 3, the change in the SdH oscillations due to the SOC is
partly responsible for the difference in MR between $B \perp I$ and $B
\parallel I$. Another factor to consider is the crossover between
classical and quantum transport regimes: As we already discussed, the
quantum regime is arrived above 4 T in the transverse configuration.
(This crossover can also be seen in the $B$ dependence of $\tan
\alpha_{\rm H}$, which is linear in $B$ in the classical regime but
saturates in the quantum regime \cite{note2}, see Fig. 1 inset.) On the
other hand, in the longitudinal configuration, the electron motion along
the current direction is {\it not} quantized, and therefore $\rho_{xx}$
for $B \parallel I$ is always ``classical". This means that in our
measurement in high magnetic fields, there is a crossover from the
classical to the quantum regime when the configuration changes from
longitudinal to transverse. Since the MR behavior is different in the
two regimes, this crossover must be partly responsible for the observed
MR anisotropy.

Although we have not been able to elucidate the mechanism for the sharp
peaks in the angular-dependent MR shown in Fig. 4(a), we can see that
there are three factors that are likely to participate in this
phenomenon: (i) the SOC which diminishes some of the peaks in the SdH
oscillations for $B \parallel I$, (ii) the static skin effect which
creates a conducting surface layer, and (iii) the crossover between
classical and quantum transport regimes. It is useful to note that the
sharp peaks weaken only gradually with increasing temperature and are
still observable at 100 K [lower inset of Fig. 4(b)], while the SdH
oscillations disappear above $\sim$20 K (lower right inset of Fig. 3);
this suggests that the sharp peak is not directly related to quantum
oscillations. We also note that the SOC is expected to affect not only
the SdH oscillations but also the static skin effect when the magnetic
field is inclined from the surface, because in such a configuration the
surface reflection of an electrons necessarily involves a transition to
a different Landau level \cite{Azbel}, and the same selection rules
imposed by the SOC as those in the SdH case \cite{Narita1} would apply.
We expect that the anomalous angular-dependent MR is a result of an
intricate interplay between the above three factors. 

In conclusion, we have observed sharp peaks in the angular-dependent MR
in PbS when the magnetic field is slightly inclined from the
longitudinal ($B \parallel I$) configuration, which is totally
unexpected for a low-carrier-density system with nearly spherical Fermi
surface. While the mechanism of this peak is to be elucidated in future,
we show that the spin-orbit coupling, the static skin effect, and the
crossover between classical and quantum transport regimes, are all
important in the magnetotransport properties of PbS. This unusual
phenomenon would help establish a general understanding of the
magnetotransport in narrow-gap semiconductors with a strong SOC, which
is important in elucidating the transport properties of topological
insulators.

\begin{acknowledgments}
This work was supported by JSPS (KAKENHI 19340078 and 2003004) and AFOSR 
(AOARD-08-4099). We thank H. D. Drew and V. Yakovenko for discussions.
\end{acknowledgments}


\begin{thebibliography}{99}
\bibitem{note1}

Analysis of the SdH data yields $T_D$ = 4 K, which gives the scattering
time 3$\times$10$^{-13}$ s. Using $m_c$ = 0.14$m_e$, we obtain $\mu_{\rm
SdH}$ = 3.8$\times$10$^{3}$ cm$^2$/Vs. The observed SdH frequency of 7.7
T corresponds to $k_F$ = 1.53$\times$10$^{6}$ cm$^{-1}$, which gives
$E_F$ = 6.4 meV, $v_F$ = 1.26$\times$10$^{7}$ cm/s, and $\ell_{\rm SdH}$
= 3.8$\times$10$^{-6}$ cm. This is contrasted to the transport
mean free path $\ell_{\rm tr}$ = 4.5$\times$10$^{-5}$ cm.
\end{thebibliography}
\end{document}